\documentclass[lettersize,journal]{IEEEtran}
\usepackage{amsmath,amssymb,amsfonts}
\usepackage{algorithmic}
\usepackage{algorithm}
\usepackage{array}
\usepackage[caption=false,font=normalsize,labelfont=sf,textfont=sf]{subfig}
\usepackage{textcomp}
\usepackage{stfloats}
\usepackage{url}
\usepackage{verbatim}
\usepackage{graphicx}
\usepackage{cite}
\usepackage{textcomp}
\usepackage{multirow}
\usepackage{booktabs}
\usepackage{adjustbox}
\begin{document}
\thispagestyle{empty}
\newpage
\onecolumn
This work has been submitted to the IEEE for possible publication. Copyright may be transferred without notice, after which this version may no longer be accessible.

\newpage
\twocolumn

\title{Distillation Learning Guided by Image Reconstruction for One-Shot Medical Image Segmentation}

\author{Feng Zhou, Yanjie Zhou, Longjie Wang, Yun Peng, David E. Carlson, and Liyun Tu
\thanks{This paragraph of the first footnote will contain the date on which you submitted your paper for review. This work was supported in part by the National Natural Science Foundation of China under Grant 62201091, the Beijing Natural Science Foundation-Haidian Original Innovation Joint Fund under Grant L232138, and STI 2030—Major Projects 2021ZD0200508. (Corresponding author: Liyun Tu)}
\thanks{Feng Zhou, Yanjie Zhou, and Liyun Tu are with the School of Artificial Intelligence, Beijing University of Posts and Telecommunications, Haidian District, Beijing 100876, China (e-mail: zhou\_feng@bupt.edu.cn; 905143096@qq.com; tuliyun@bupt.edu.cn). }
\thanks{Longjie Wang is with the Department of Orthopaedics, Peking University Third Hospital, Haidian District, Beijing 100191, China (e-mail: wlj2096@163.com).}
\thanks{Yun Peng is with the Department of Radiology, MOE Key Laboratory of Major Diseases in Children, Beijing Children’s Hospital, Capital Medical University, National Center for Children’s Health, Xicheng District, Beijing 100045, China (e-mail: ppengyun@hotmail.com).}
\thanks{David E. Carlson is with the Department of Biostatistics and Bioinformatics and the Department of Civil and Environmental Engineering, Duke University, Durham, NC 27708, USA (e-mail: david.carlson@duke.edu).}}

\markboth{Journal of \LaTeX\ Class Files,~Vol.~14, No.~8, August~2021}%
{Shell \MakeLowercase{\textit{et al.}}: A Sample Article Using IEEEtran.cls for IEEE Journals}

\IEEEpubid{\makebox[\columnwidth]{0000--0000/00\$00.00~\copyright~2021 IEEE\hfill}}
\IEEEpubidadjcol

\maketitle

\begin{abstract}
Traditional one-shot medical image segmentation (MIS) methods use registration networks to propagate labels from a reference atlas or rely on comprehensive sampling strategies to generate synthetic labeled data for training. However, these methods often struggle with registration errors and low-quality synthetic images, leading to poor performance and generalization. To overcome this, we introduce a novel one-shot MIS framework based on knowledge distillation, which allows the network to directly 'see' real images through a distillation process guided by image reconstruction. It focuses on anatomical structures in a single labeled image and a few unlabeled ones. A registration-based data augmentation network creates realistic, labeled samples, while a feature distillation module helps the student network learn segmentation from these samples, guided by the teacher network. During inference, the streamlined student network accurately segments new images. Evaluations on three public datasets (OASIS for T1 brain MRI, BCV for abdomen CT, and VerSe for vertebrae CT) show superior segmentation performance and generalization across different medical image datasets and modalities compared to leading methods. Our code is available at https://github.com/NoviceFodder/OS-MedSeg.
\end{abstract}


\section{Introduction}

\begin{figure}[t]
    \centering
    \includegraphics[width=0.48\textwidth]{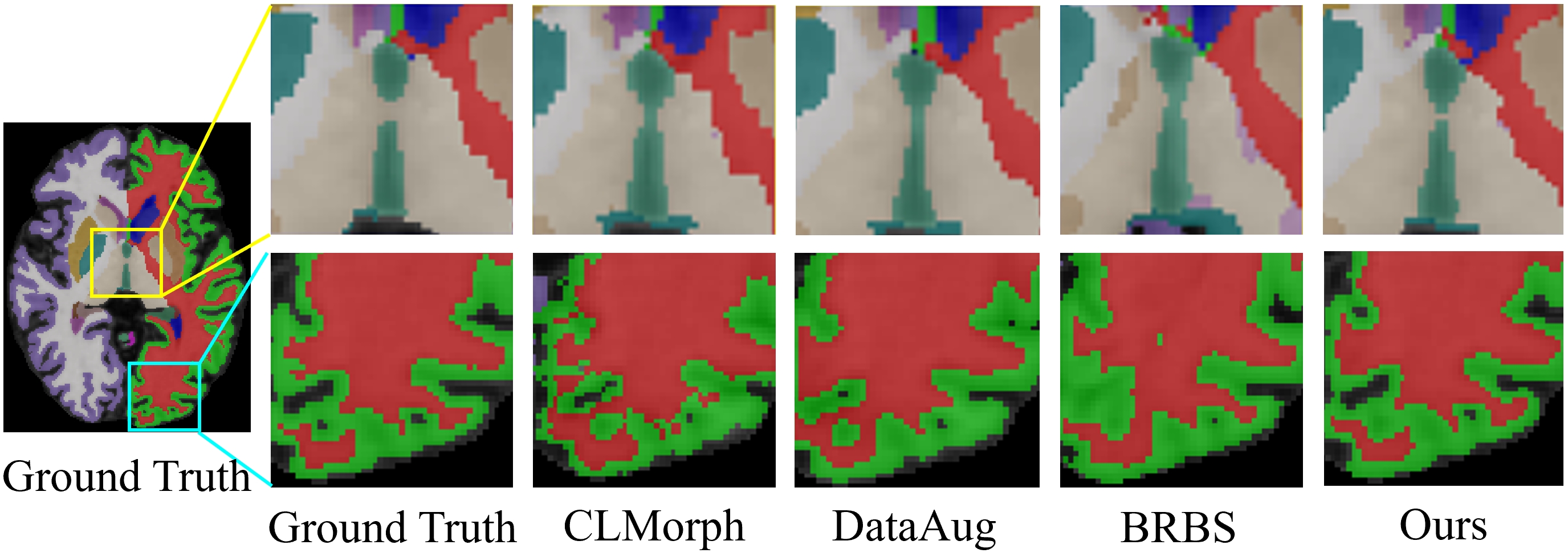}
    \caption{Overview of our problem. Our proposed method achieves natural, realistic, and smooth segmentation, outperforming current state-of-the-art one-shot methods (CLMorph \cite{liu2023contrastive}, DataAug \cite{zhao2019data}, and BRBS \cite{he2022learning}).    
    }
    \label{fig:intro}
\end{figure}

Segmentation is a fundamental task in medical imaging analysis, involving the identification and delineation of specific anatomical structures or regions of interest in various medical images, such as Computed Tomography (CT) and Magnetic 
Resonance Imaging (MRI). Precise medical image segmentation (MIS) is crucial for improving disease diagnosis \cite{Ma2024}, aiding treatment planning \cite{Isensee2021}, monitoring disease progression \cite{Ouyang2020}, and facilitating patient management \cite{Antonelli2022}. 

Existing methods are generally specialized for particular tasks or built upon foundational segmentation models \cite{Minaee2022}. Examples of these foundational models include the Segment Anything Model (SAM) \cite{kirillov2023segany} and segment everything everywhere with multi-modal prompts \cite{zou2023segment}, which aim to create universal models. While these models demonstrate impressive adaptability and performance across different tasks, they typically rely on extensive labeled datasets to achieve high accuracy. The manual process of labeling anatomical structures and pathological regions in 3D medical images is exhaustive, time-consuming, and requires expertise, posing challenges for training fully or semi-supervised segmentation methods. Consequently, self-supervised \cite{wang2019panet} and few-shot \cite{grill2020bootstrap,wang2020self} segmentation techniques are being explored to overcome data availability limitations and reduce reliance on extensive, well-representative annotations.

One-shot MIS methods, a subset of few-shot learning, commonly employ registration networks \cite{balakrishnan2019voxelmorph,chen2022transmorph} to align a well-labeled atlas with unlabeled images for label prediction or leverage synthetic labeled data to train segmentation networks. These methods encounter difficulties such as voxel intensity variations, which challenge the spatial transformer's ability to accurately align two images \cite{liu2023contrastive}. To enhance training stability, registration-based models often incorporate techniques like forward-backward consistency between atlas and target images \cite{wang2020lt}, bi-directional spatial transformations for inverse consistency \cite{Zheng2022InverseConsistency}, or multi-scale and cascaded Convolutional Neural Network (CNN) models to decompose deformation fields \cite{kang2022dual,zhao2019unsupervised,lv2022joint}. Despite achieving impressive registration results, atlas-based segmentation can suffer from blurring of image details due to its dependence on image similarity. Models trained on generated images share similarities with DataAug \cite{zhao2019data}, which often prioritize high-quality synthetic datasets and lack the anatomical guidance needed to capture tissue details in real images. This affects their ability to generalize across different datasets and imaging modalities due to the heterogeneity inherent in medical imaging (Fig. \ref{fig:intro}).

In this work, we present a novel one-shot MIS method for precise tissue segmentation using a reconstruction-guided distillation learning framework. Our main contributions are:

\begin{itemize}
  
  \item {\textit{We develop an innovative optimization strategy that effectively captures the details of anatomical structures from unlabeled images to guide one-shot medical image segmentation}. Unlike conventional methods hindered by synthetic image quality and registration errors, our approach leverages real image information from limited unlabeled data to facilitate feature representation learning.}


  \item {\textit{We introduce a novel distillation learning framework that transfers reconstruction features to the segmentation learning process}. The framework enables the student network to use prior knowledge from the teacher network for accurate segmentation. Additionally, we incorporate a novel cosine similarity loss to achieve smoother and more natural segmentation.}


  \item{\textit{Our method offers a streamlined, lightweight inference for unseen images through a simplified student network}. It consistently outperforms existing state-of-the-art (SOTA) one-shot segmentation methods and generalizes across multiple datasets (brain, abdomen, and vertebrae) with different modalities (MRI and CT).}
  

\end{itemize}

\section{Related Work}

\subsection{One-Shot Medical Image Segmentation}
\subsubsection{Atlas-Based Segmentation (ABS)} A well-established one-shot MIS paradigm that uses the relationship between segmentation labels and images in atlas-label pairs. It involves registering labeled images (atlases) to unlabeled images to create an indirect mapping for segmentation results \cite{iglesias2015multi}. Traditional ABS methods often rely on conventional registration techniques in software like ANTs \cite{avants2009advanced} and FreeSurfer \cite{fischl2012freesurfer}, including rigid, affine \cite{arun1987least}, and SyN \cite{avants2008symmetric} registrations, which are time-consuming and limited by image modality. Recent advancements in deep learning (DL) have led to unsupervised registration models that enable faster and more accurate alignments between atlases and target images \cite{balakrishnan2018unsupervised}. However, these models often suffer from registration errors, leading to inaccurate label transfer. Approaches to mitigate these errors include probabilistic generative models \cite{dalca2019unsupervised}, hybrid Transformer-ConvNet models \cite{chen2022transmorph}, cycle-correspondence \cite{wang2020lt}, and contrastive learning \cite{liu2023contrastive}. Despite these efforts, performance remains constrained by the similarity between atlas and target images, and robustness issues arise in tasks with significant deformations (e.g., abdominal CT) \cite{van20213d}, limiting their effectiveness for large organ segmentation.

\subsubsection{Learning Registration to Learn Segmentation (LRLS)} A recent one-shot MIS paradigm \cite{he2020deep} that employs registration networks to learn voxel-wise correspondences between labeled and unlabeled data, creating labeled pseudo-datasets for segmentation. DataAug \cite{zhao2019data} advanced this paradigm by introducing various spatial and appearance transformations to enhance synthetic data diversity. Ding et al. \cite{ding2021modeling} improved on DataAug by using a VAE to generate varied pseudo-datasets from a continuous latent space. However, these methods do not address registration errors, which can mislead subsequent segmentation learning. Methods like DeepAtlas \cite{xu2019deepatlas}, DeepRS \cite{he2020deep}, and BRBS \cite{he2022learning} optimize registration and segmentation networks jointly. The registration network generates synthetic labels for the segmentation network, which then refines the registration network. Jiang et al. \cite{jiang2022one} applied this joint optimization framework in longitudinal thoracic cone beam CT segmentation. However, these methods do not fully utilize real unlabeled images, which have clearer anatomical structure information compared to synthetic images. Joint training also presents challenges in parameter volume and model convergence. Their adaptability to abdominal organ and vertebrae segmentation and robustness across different imaging modalities remain underexplored.

\subsection{Distillation Learning}
Distillation learning, originally developed for object classification \cite{hinton2015distilling}, involves creating simpler models (students) from a complex, pre-trained teacher network. This process entails regularizing the student network to mimic the teacher's probabilistic outputs or its intermediate features \cite{yim2017gift}. It has been effectively applied in natural image segmentation, and recently in medical image analysis, particularly for uni-modal \cite{Chen2022} and cross-modality \cite{Dou2020,Jiang2022,LiYWH20} lesion segmentation. A significant aspect of this approach is the need for a high-capacity teacher network, pre-trained on extensive data, especially when real-time analysis is computationally challenging.

A distinct distillation learning approach, known as collaborative learning \cite{song2018collaborative}, involves training multiple weak learners collaboratively on the same task without needing a large pre-trained teacher network, thereby enhancing robustness and accuracy through diverse parameter initialization and representation extraction \cite{furlanello2018born}. Here, a teacher's knowledge can be refined through self-training of seeded student networks of similar architectural complexity. This approach focuses on robustness to initial conditions rather than imaging conditions \cite{Yang2019}. Knowledge augmentation in this context also involves leveraging different information sources and additional datasets for training \cite{Dou2020,LiYWH20}, with regularization achieved by aligning the student's features with the teacher's \cite{Chen2022}. Unlike these methods, our approach uses the teacher model to help the less informative student model extract task-relevant features.


\section{Methods}

\begin{figure*}[t]
     \centering
     \includegraphics[width=0.95\textwidth]{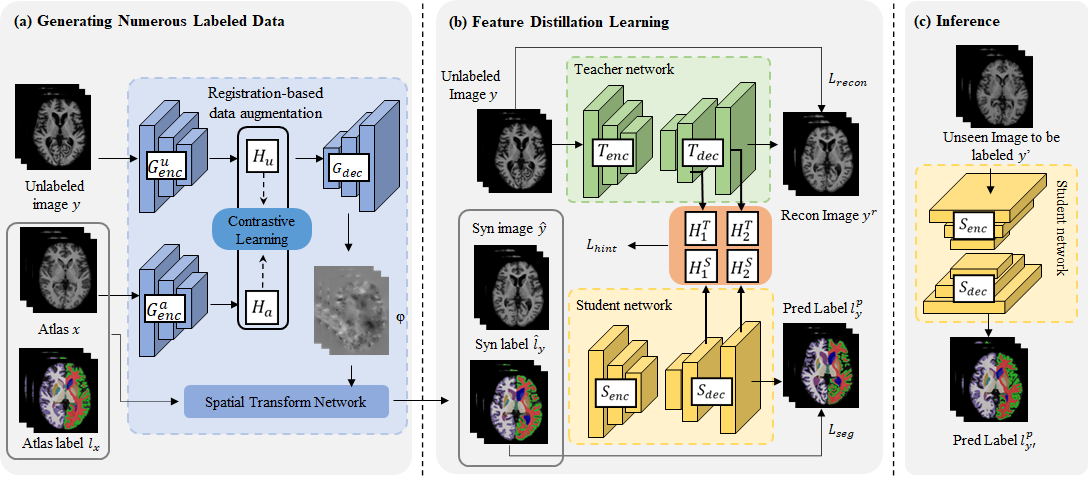}
     \caption{Schematic of our one-shot medical image segmentation framework, consisting of three stages: (a) \textbf{Generating Labeled Data} using image registration and contrastive learning, (b) \textbf{Feature Distillation Learning} through image reconstruction and segmentation, and (c) \textbf{Inference} with a lightweight student network predicting segmentation labels on unknown images. $G$ denotes the registration network, $T$ and $S$ are the teacher and student networks, and $H$ represents the extracted features.}     
     \label{fig:Framework}
 \end{figure*}


\subsection{Data Augmentation with Image Registration and Contrastive Learning}
Our proposed framework (Fig. \ref{fig:Framework}) begins with a registration-based data augmentation network to generate well-labeled training samples approximating real images. Inspired by CLMorph \cite{liu2023contrastive}, a SOTA unsupervised registration method that integrates contrastive learning for image-to-image deformation, we designed and implemented a data augmentation network (Fig. \ref{fig:Framework}a) to capture the distributions of transformations between an atlas and each unlabeled image.

\subsubsection{Registration-based Data Augmentation Network} In this work, we use a variant of CLMorph. Unlike the original model (CLMorph \cite{liu2023contrastive}), we modified the encoder network to eliminate the reparameterization trick for Gaussian distribution, simplifying the model by directly encoding the input image without introducing random variability in the latent space. We doubled the number of convolution filters to enhance feature extraction efficiency. Additionally, we expanded the original model's similarity loss and smooth loss to better handle the intensity and deformation variability across datasets of different organs and modalities (see Section \ref{sec:opt-of-reg} for more details). Note that the registration network is not the focus of this work, as our framework adapts to any unsupervised registration network.


Consider $Y=\{y^i\}_{i=1}^N$ as a collection of $N$ images and the pair $(x,l_{x})$ as an atlas along with its corresponding segmentation label. Here, $y^i$, $x$, and $l_x$ are defined over an $n$-dimensional spatial domain $\Omega\subset\mathbb{R}^n$, where $n = 3$ is used throughout this work. Given an atlas image $x$ and an unlabeled image $y^{i}$ as the input, the registration network is trained to learn 3D image-to-image alignment maps. The network employs two weight-shared CNN encoders, $G^{u}_{enc}$ and $G^{a}_{enc}$, to extract the highly semantic features from the unlabeled and atlas images. We then use a decoder $G_{dec}$ to integrate the feature maps output by the two encoders. The specific operation involves concatenating the feature maps from the two CNN outputs. The decoder employs skip connections, recursively utilizing high-level semantic information from feature maps to extract features with low-level detail information until feature maps match the resolution of the input image and produces the corresponding deformation field $\phi$. We create a labeled synthetic example $(\hat{y}^{i},\hat{l_y}^{i})$ by applying the transformations computed from the labeled atlas to the target volumes:

\begin{equation}
\begin{aligned}
\hat{y}^{i} = x \circ \phi,\\
\hat{l_y}^{i} = l_x \circ \phi,
\end{aligned}
\end{equation}
where $\circ$ denotes a warping operation facilitated by spatial transformer networks \cite{jaderberg2015spatial}. The newly labeled training examples are subsequently integrated into the labeled training set for a supervised segmentation network.

\subsubsection{Feature-Level Contrastive Learning} 
We used contrastive learning to extract features with rich information to improve the registration performance. Formally, given a set of images $Y=\{y^i\}_{i=1}^N$, we treat $(x, y^{i})$ as an augmented image pair, and other images in $Y$ as negative samples. Moreover, we denote $sim(u,v)=\frac{u^\mathrm{T}v}{||u||\cdot||v||}$ as the cosine similarity between $u$ and $v$. We formulate the contrastive loss function as follows:
\begin{equation}
\label{contrastive-loss}
\begin{aligned}
&\mathcal{L}_{contrast}(H_u, H_a)=  \\
& -\log\frac{\exp(\operatorname{sim}(H_u, H_a)/\tau)}{\sum_{i \in N} \textbf{1}_{i \neq y^{i}} \exp(\operatorname{sim}(H_u, H_a)/\tau)},
\end{aligned}
\end{equation}
where $H_u$ and $H_a$ denote the generated features from the CNN encoder,  which are utilized to maximize the consistency between images and enhance the authenticity of generated images. The indicator $\textbf{1}_{i\neq y^{i}}\in\{0,1\}$ takes the value 1 only when $i\neq y^{i}$. $\tau$ is the temperature hyperparameter \cite{wu2018unsupervised}.

\subsection{Feature Distillation Learning with Image Reconstruction for Segmentation}
In Fig. \ref{fig:Framework}b, we adopt a teacher-student network architecture for feature distillation learning. Each component of this architecture will be thoroughly explained in this section.
\begin{figure}[t]
    \centering    \includegraphics[width=0.48\textwidth]{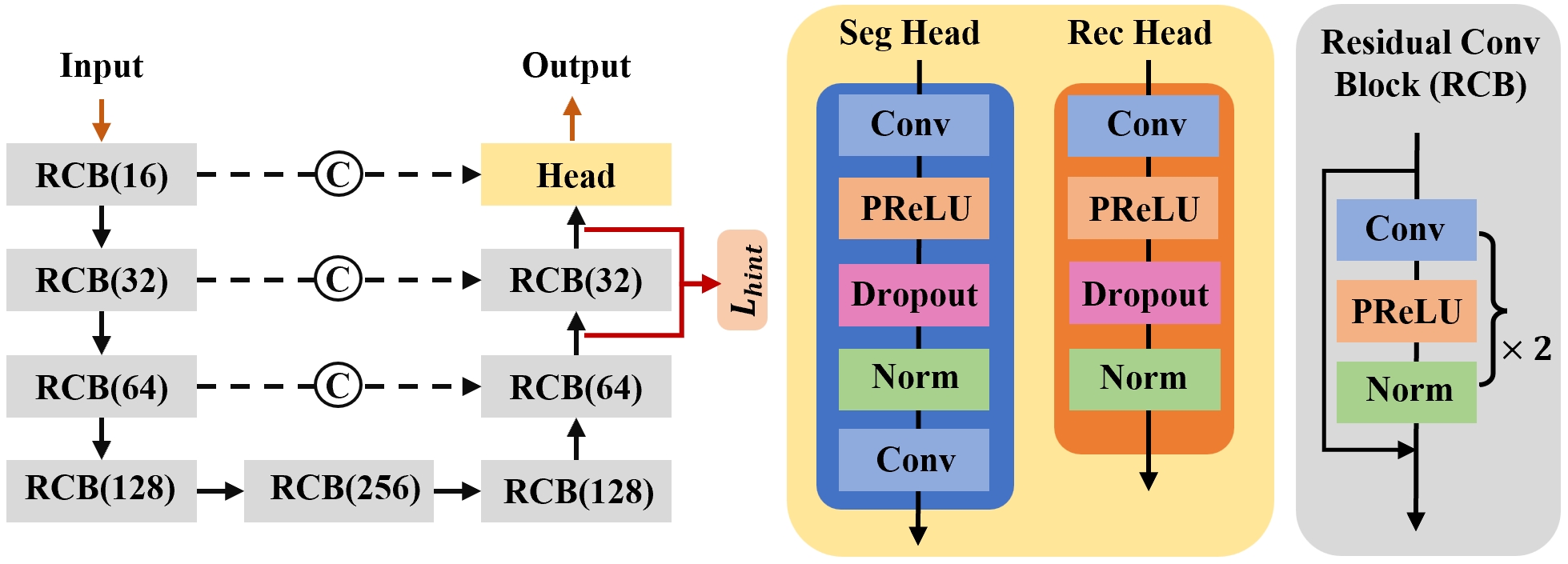}
    \caption{The proposed teacher-student network uses a residual join U-Net, with each residual block having convolution, PReLU, and layer normalization. It has two output headers: Seg Head for segmentation (student network) and Rec Head for reconstruction (teacher network). ${L}_{hint}$ is calculated from the last two layers of both networks.}  
    \label{fig:unet-detailed}
\end{figure}

\subsubsection{Teacher-Student Network}
Given an input image $y^{i}$, the teacher network aims to acquire knowledge about anatomical features inherent in real images by means of the reconstruction process. For a set of paired data $\{(\hat{y}^{i},\hat{l_y}^{i})\}$, the student network focuses on capturing unique features within the regions of interest in the synthetic images and their corresponding labels. The paired networks are both based on the 3D U-Net architecture, incorporating residual connections \cite{he2016deep} within the encoder and output paths (Fig. \ref{fig:unet-detailed}). Despite sharing a common structure, each network is designed with distinct output heads.

\subsubsection{Feature Distillation Learning} This module is designed to guide the student network in segmenting synthetic images with the assistance of the reconstruction features derived from the teacher network. It is noteworthy that the synthetic images $\{(\hat{y}^{i},\hat{l_y}^{i})\}$ provided to the student network are registered from atlas $x$ to unlabeled images $y^{i}$. Therefore, the input images of the two networks correspond to each other. The reconstruction process of the teacher network effectively learns the anatomical features of the real images $y^{i}$, compensating for errors introduced during the registration process that might lead to unrealistic synthetic images. This, in turn, enhances the segmentation performance of the student network.

Previous studies indicate that features closer to the output are highly correlated with the output task \cite{lin2017refinenet}. Based on this, we experimentally analyzes the number of feature layers used for distillation learning, calculating hint loss for features from the last 1-5 layers of both teacher and student networks. As detailed in Section \ref{hyper-impact}, we found that selecting the last two layers yielded the best Dice coefficient for the results reported in this work. In Section \ref{DL-impact}, we further studied the impact of different distillation losses on our framework. Instead of using the commonly used L2 norm \cite{yim2017gift}, we employ cosine similarity to decrease the feature distance, thereby enhancing the student network's performance with the following loss:

\begin{equation}
\label{hint-loss}
\mathcal{L}_{hint}(\phi_{C_i},\phi_{M_i})=\sum_{i=1}^N(1-\text{cos}(\phi_{C_i},\phi_{M_i})),\end{equation}
\begin{equation}\text{cos}(\phi_{C_i},\phi_{M_i})=\frac{\phi_{C_i}\cdot\phi_{M_i}}{\|\phi_{C_i}\|\cdot\|\phi_{M_i}\|},\end{equation}
where $\phi_{C_i}$, $\phi_{M_i}$ are the $i_{th}$ layer features computed from the two networks, and $N$ is the total number of features.

Cosine similarity loss emphasizes angular relationships in probability distributions, focusing on relative ranking relationships crucial for learning semantic information and anatomical features in images. This angular information is particularly significant in segmentation tasks, aiding understanding of semantic structures, boundaries, and contours. Additionally, cosine similarity loss exhibits robustness to noise, facilitating effective learning from actual image data. Given the task framework and the benefits of cosine similarity loss, it proves more suitable for transferring prior knowledge (e.g., the learnt image anatomical features) from the teacher network to guide superior performance in image segmentation tasks.


\subsection{Optimization and Inference Stage}
\subsubsection{Optimization of Registration-based Data Augmentation}
\label{sec:opt-of-reg}
The objective function of the registration-based network, $\mathcal{L}_{reg}$, consists of three components: a similarity loss $\mathcal{L}_{sim}$ to penalize differences in appearance, a deformation smoothness regularization $\mathcal{L}_{smooth}$ to penalize local spatial variations in $\phi$, and a contrastive loss $\mathcal{L}_{contrast}$ (Eq. \eqref{contrastive-loss}) to penalize incorrect image-to-image alignment:
\begin{equation}
\label{syn-loss}
\mathcal{L}_{reg} = \mathcal{L}_{sim} + \alpha \mathcal{L}_{smooth} + \beta \mathcal{L}_{contrast},
\end{equation}
where $\alpha$ and $\beta$ are the hyper-parameters balancing these three components.

\textbf{Image Similarity Measure.} We used two widely-used similarity metrics for $\mathcal{L}_{sim}$ to better handle the registration tasks across different modalities and organs. The first, local cross-correlation (CC) loss $\mathcal{L}_{CC}$, was used for the OASIS and VerSe datasets. This metric is robust to intensity variations and captures details in homogeneous regions \cite{balakrishnan2019voxelmorph}, making it suitable for brain tissue and vertebrae image registration tasks. The $\mathcal{L}_{CC}$ is formulated as:

\begin{equation}
\label{ncc-loss}
\begin{aligned}
&\mathcal{L}_{CC}(I_f, I_w) = \\
&\sum_{\mathbf{p} \in \Omega} \frac{\left(\sum_{\mathbf{p}_i} \big(I_{f}(\mathbf{p}_i)-\bar{I}_{f}(\mathbf{p})\big)\big(I_{w}(\mathbf{p}_i)-\bar{I}_{w}(\mathbf{p})\big)\right)^2}{\left(\sum_{\mathbf{p}_i} \big(I_{f}(\mathbf{p}_i)-\bar{I}_{f}(\mathbf{p})\big)^2\right)\left(\sum_{\mathbf{p}_i} \big(I_{w}(\mathbf{p}_i)-\bar{I}_{w}(\mathbf{p})\big)^2\right)},
\end{aligned}
\end{equation}
where $I_f$ and $I_w$ represent $y^{i}$ and $x \circ \phi$ respectively, $\bar{I}_{f}(\mathbf{p})$ and $\bar{I}_{w}(\mathbf{p})$ denote the mean voxel value within a local window of size $n^3$ centered at voxel $\mathbf{p}$. We use $n=9$ as recommended by \cite{balakrishnan2018unsupervised}. The second, mutual information (MI) loss $\mathcal{L}_{MI}$, was used for the BCV dataset. MI does not rely on specific intensity values but utilizes the statistical information between images, making it effective for handling complex, nonlinear transformations, such as those encountered in abdominal organ registration tasks \cite{pluim2003mutual}. $\mathcal{L}_{MI}$ is formulated as:

\begin{equation}
\label{mi-loss}
\begin{aligned}
&\mathcal{L}_{MI}(I_f, I_w) = \sum_{i,j} p(I_f=i, I_w=j) \log \frac{p(I_f=i, I_w=j)}{p(I_f=i) p(I_w=j)},
\end{aligned}
\end{equation}
where $p(I_f=i, I_w=j)$ is the joint probability distribution of $I_f$ and $I_w$, $p(I_f=i)$ and $p(I_w=j)$ are the marginal probability distributions of  $I_f$ and $I_w$, respectively.

\textbf{Deformation Field Regularization.} We also used two different regularizers as $\mathcal{L}_{smooth}$ to better handle the deformation field regularization. The first was the diffusion regularizer \cite{balakrishnan2019voxelmorph}, which is employed in OASIS and VerSe datasets:

\begin{equation}
\label{grad-loss}
\mathcal{L}_{diffusion}(\phi)=\sum_{\mathbf{p} \in \Omega}\|\nabla \mathbf{u}(\mathbf{p})\|^2,
\end{equation}
where the $\nabla \mathbf{u}(\mathbf{p})$ represents the spatial gradients of the displacement field $\mathbf{u}$. These gradients are approximated using forward differences: $\frac{\partial\mathbf{u}(\mathbf{p})}{\partial x} \approx \mathbf{u}(p_{x}+1, p_{y}, p_{z}) - \mathbf{u}(p_{x}, p_{y}, p_{z})$, with similar approximations for $\frac{\partial\mathbf{u}(\mathbf{p})}{\partial y}$ and $\frac{\partial\mathbf{u}(\mathbf{p})}{\partial z}$.

The second regularizer used for the BCV dataset was the bending energy regularizer \cite{rueckert1999nonrigid}, which penalizes sharply curved deformations and is particularly useful for abdominal organ registration \cite{chen2022transmorph}. Bending energy operates on the second derivative of the displacement field $\mathbf{u}$, defined as:

\begin{equation}
\label{bending energy-loss}
\begin{aligned}
&\mathcal{L}_{bending}(\phi) = \sum_{p \in \Omega} \|\nabla^2 \mathbf{u}(p)\|^2 \\
&= \sum_{p \in \Omega} \Bigg[
\left( \frac{\partial^2 \mathbf{u}(p)}{\partial x^2} \right)^2 + 
\left( \frac{\partial^2 \mathbf{u}(p)}{\partial y^2} \right)^2 + 
\left( \frac{\partial^2 \mathbf{u}(p)}{\partial z^2} \right)^2 + \\
&\quad 2 \left( \frac{\partial^2 \mathbf{u}(p)}{\partial xz} \right)^2 + 
2 \left( \frac{\partial^2 \mathbf{u}(p)}{\partial xy} \right)^2 + 
2 \left( \frac{\partial^2 \mathbf{u}(p)}{\partial yz} \right)^2 
\Bigg],
\end{aligned}
\end{equation}
where the derivatives were estimated using the same forward
differences that were used previously.

\subsubsection{Optimization of Feature Distillation Learning}
For the teacher network, we compute the mean squared error (MSE) similarity between the real image $y$ and its reconstructed counterpart $y^{r}$ as the reconstruction loss, aiming to enhance the overall reconstruction performance, as expressed in the following:
\begin{equation}
\label{recon_loss}
\mathcal{L}_{recon}(y,y^{r})=\frac{1}{|\Omega|}\sum_{p\in\Omega}\left[y(\mathbf{p})-y^{r}(\mathbf{p})\right]^{2}.
\end{equation}

In the case of the student network, we compute the segmentation loss through the cross-entropy function to minimize the difference between the predicted label $l^{p}$ and the input synthetic label $\hat{l}$, as formulated in Eq. \eqref{seg-loss}. 
\begin{equation}
\label{seg-loss}
\mathcal{L}_{seg}(\hat{l}, l^{p}) = -\sum_{c=1}^C \hat{l}_c \log l^{p}_c,
\end{equation}
where $C$ represents the number of classes.

The objective function of feature distillation learning $\mathcal{L}_{kd}$ is expressed as:
\begin{equation}
\label{kd-loss}
\mathcal{L}_{kd} = \mathcal{L}_{seg} + \lambda_{recon} \mathcal{L}_{recon}  + \lambda_{hint} \mathcal{L}_{hint},
\end{equation}
where $\lambda_{recon}$ and $\lambda_{hint}$ are the weighting
coefficients for $\mathcal{L}_{recon}$ (Eq. \eqref{recon_loss}) and $\mathcal{L}_{hint}$ (Eq. \eqref{hint-loss}), respectively.

\subsubsection{Inference}

In the final inference process, we retain only the well-trained lightweight student network (Fig. \ref{fig:Framework}c). This student network integrates region-of-interest segmentation features extracted from synthetic data and anatomical features extracted from real images in the teacher network. This integration enables precise segmentation of unknown images.
\section{Experiments and Results}
\begin{table*}[th]
\scriptsize
\setlength{\tabcolsep}{2pt}
    \centering
    \caption{Quantitative evaluation on the OASIS, BCV, and VerSe datasets used the Dice Similarity Coefficient (DSC) and the 95th percentile Hausdorff Distance (HD$_{95}$mm) metrics. The best result, excluding the upper limit, is highlighted in bold. 
    }
    \label{tab:three datasets res}
    \resizebox{0.85\textwidth}{!}{
    \begin{tabular}{lcccccccc}
        \toprule
        \multirow{3}{*}{\textbf{Method}} & \multirow{3}{1.5cm}{\centering \textbf{Type}} & \multicolumn{2}{c}{\textbf{OASIS}} & \multicolumn{2}{c}{\textbf{BCV}} & \multicolumn{2}{c}{\textbf{VerSe}} \\
        \cmidrule(lr){3-4} 
         \cmidrule(lr){5-6} \cmidrule(lr){7-8} 
        & & \textbf{DSC $\uparrow$} & \textbf{HD$_{95}mm\downarrow$} & \textbf{DSC $\uparrow$} &\textbf{ HD$_{95}mm\downarrow$} & \textbf{DSC $\uparrow$} & \textbf{HD$_{95}mm\downarrow$} \\
        \midrule  
        Fully supervised \cite{he2016deep} & Upper bound 
        & \textit{0.906$\pm$0.014} & \textit{1.346$\pm$0.380}
        & \textit{0.919$\pm$0.056} & \textit{4.932$\pm$4.652}
        & \textit{0.964$\pm$0.018} & \textit{1.302$\pm$1.250}\\
         \midrule 
        U-Net \cite{ronneberger2015u} & LS 
        & 0.684$\pm$0.064 & 14.373$\pm$7.161
        & 0.538$\pm$0.110 & 68.787$\pm$7.869
        & 0.283$\pm$0.078 & 56.783$\pm$18.764 \\
        ResUNet \cite{diakogiannis2020resunet} & LS 
        & 0.724$\pm$0.048 & 8.237$\pm$3.209
        & 0.609$\pm$0.148 & 48.744$\pm$7.207
        & 0.465$\pm$0.094 & 28.691$\pm$10.867 \\
        \midrule
        Rigid \cite{arun1987least} & Trad 
        & 0.597$\pm$0.049 & 4.163$\pm$0.709 
        & - & -
        & 0.532$\pm$0.140 & 3.689$\pm$1.354 \\
        Affine \cite{arun1987least} & Trad 
        & 0.601$\pm$0.050 & 4.131$\pm$0.719 & 0.561$\pm$0.101 & 13.569$\pm$4.009 &  0.633$\pm$0.112 & 3.870$\pm$2.303 \\
        SyN \cite{avants2008symmetric} & Trad
        & 0.784$\pm$0.021 & 2.627$\pm$0.438 & 0.768$\pm$0.108 & 11.914$\pm$5.014 & 0.830$\pm$0.101 & 1.794 $\pm$1.417 \\
        \midrule
        VoxelMorph-1 \cite{balakrishnan2018unsupervised} & ABS    
        & 0.793$\pm$0.019 & 3.085$\pm$0.531 
        & 0.700$\pm$0.097 & 13.271$\pm$2.363
        & 0.770$\pm$0.136 & 2.223$\pm$0.217  \\
        VoxelMorph-2 \cite{balakrishnan2019voxelmorph} & ABS 
        & 0.800$\pm$0.017 & 3.220$\pm$0.574 
        & 0.726$\pm$0.103 & 12.917$\pm$1.679
        & 0.790$\pm$0.136 & 2.113$\pm$0.203 \\
        TransMorph \cite{chen2022transmorph} & ABS 
        & 0.796$\pm$0.017 & 2.654$\pm$0.396
        & 0.766$\pm$0.103 & 12.908$\pm$4.007
        & 0.817$\pm$0.132 & 2.667$\pm$2.563\\
        CLMorph \cite{liu2023contrastive} & ABS 
        & 0.799$\pm$0.019 & 3.214$\pm$0.555  
        & 0.726$\pm$0.108 & 12.802$\pm$1.257 
        & 0.803$\pm$0.127 & 2.176$\pm$0.243\\
        \midrule
        DataAug \cite{zhao2019data} & LRLS 
        & 0.823$\pm$0.018 & 2.732$\pm$0.500
        & 0.822$\pm$0.103 & 11.006$\pm$6.061
        & 0.868$\pm$0.059 & 2.593$\pm$1.216\\
        BRBS \cite{he2022learning} & LRLS 
        & 0.811$\pm$0.021 & 3.099$\pm$0.546
        & 0.806$\pm$0.113 & 14.282$\pm$5.638
        & 0.892$\pm$0.077 & 2.754$\pm$0.807\\
        \midrule
        \textbf{Ours} & LRLS 
        & \textbf{{0.854$\pm$0.023}} & \textbf{{1.716$\pm$0.252}}
        & \textbf{{0.846$\pm$0.084}} & \textbf{{10.779$\pm$2.579}}
        & \textbf{{0.924$\pm$0.028}}& 
        \textbf{{1.593$\pm$1.768}}
        \\

        \bottomrule
    \end{tabular}}
\end{table*}

\subsection{Datasets and Preprocessing}
Three public datasets were used in this work.

1) Open Access Series of Imaging Studies (\textbf{OASIS}): contains 414 T1-weighted MRI images provided by \cite{hoopes2022learning, marcus2007open}, including 1 atlas, 334 training, and 79 test volumes with labels for 35 anatomical structures. Standard pre-processing with FreeSurfer \cite{fischl1999cortical} and SAMSEG \cite{puonti2016fast} included skull stripping, bias correction, registration, and resampling to FreeSurfer's Talairach space. The images were cropped to $160\times192\times224$ and resampled to an isotropic voxel size of $1 mm$.

2) Multi-Atlas Labeling Beyond the Cranial Vault (\textbf{BCV}): includes 50 abdomen CT scans from learn2reg2021 \cite{hering2022learn2reg}, containing 1 atlas, 29 training, and 20 test images. Pre-processing includes affine pre-alignment, cropping, padding, and resampling to $192 \times 160 \times 192$ with $2 mm$ isotropic voxels in size. Annotations are provided for the liver, spleen, right kidney, and left kidney.

3) Large Scale Vertebrae Segmentation Challenge (\textbf{VerSe}): a subset of the MICCAI VerSe20 Challenge dataset \cite{sekuboyina2021verse}, containing 86 thoracic vertebrae CT images: 1 atlas, 69 training, and 16 test images with labels for 13 anatomical structures. Pre-processing with ANTs \cite{avants2009advanced} includes rigid pre-alignment, cropping, padding, resampling, and windowing, resulting in a resolution of 64 × 96 × 192 with $2 mm$ isotropic voxels.


All atlas images were selected based on the highest image-level similarity to the test set from the training data \cite{zhao2019data,balakrishnan2019voxelmorph}. Similarity was calculated by computing the Normalized Cross-Correlation (NCC) score between the atlas and each test image, then averaging the scores.

\subsection{Experimental Settings}
\label{Experimental_Settings}
We evaluate our method using a fully supervised 3D U-Net with residual connections, trained on all dataset labels as the upper bound, and compare it against 11 widely used one-shot MIS methods categorized into 4 groups.

1) \textbf{LS}: We used data augmentation (random rotation, shear, translation, scaling) to directly learn segmentation frameworks. U-Net  \cite{ronneberger2015u} and ResUNet \cite{he2016deep} established baseline performance for supervised MIS in one-shot scenarios.

2) \textbf{Trad}: Traditional atlas-based one-shot MIS methods using Advanced Normalization Tools (ANTs) \cite{avants2009advanced}, such as rigid, affine \cite{arun1987least}, and symmetric normalization \cite{avants2008symmetric}. Although considered SOTA for classical intensity-based registration, these methods are time-consuming and offer limited accuracy and flexibility.

3) \textbf{ABS}: Atlas-based one-shot deep learning MIS methods, including VoxelMorph \cite{balakrishnan2018unsupervised,balakrishnan2019voxelmorph}, Transmorph \cite{chen2022transmorph}, and CLMorph \cite{liu2023contrastive}, are compared to highlight limitations due to registration errors and dissimilarity between atlas and target images. For VoxelMorph, we used two variants: VoxelMorph-1 and VoxelMorph-2, with the latter doubling the convolution filters of the former.


4) \textbf{LRLS}: Contains two SOTA learning registration to learn segmentation methods, DataAug \cite{zhao2019data} and BRBS \cite{he2022learning}.

For a fair comparison, all ABS and LRLS methods use the same similarity and smooth loss as our model. Each LRLS method employs CLMorph as the synthesis network and a 3D U-Net with residual connections as the segmentation network to avoid interference from network architecture differences.

\begin{figure}[ht]
    \centering
    \includegraphics[width=0.35\textwidth]{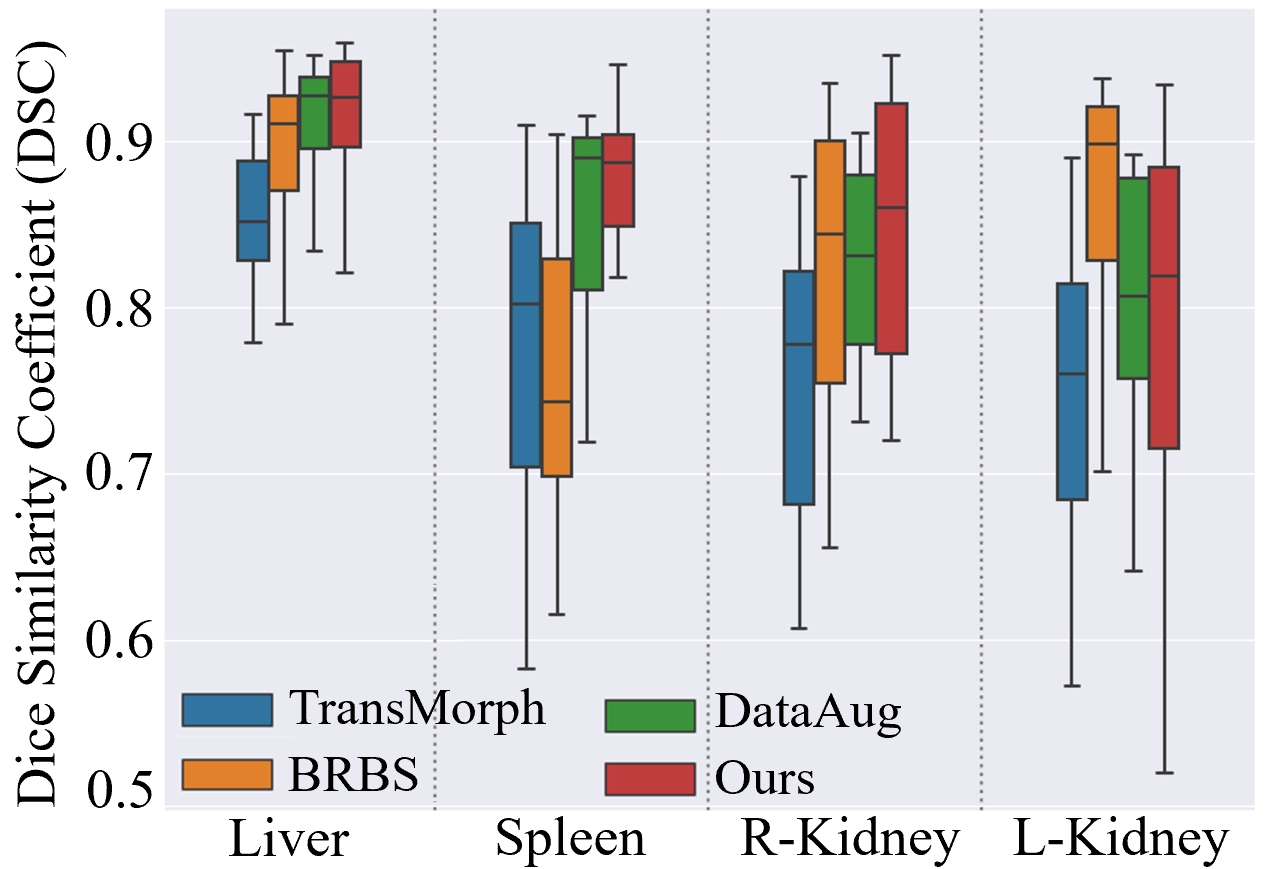}
    \caption{Comparisons with SOTA one-shot MIS methods on BCV.}
    \label{fig:bcv_boxplots}
\end{figure}

\begin{figure*}[ht]
    \centering
    \includegraphics[width=0.9\textwidth]{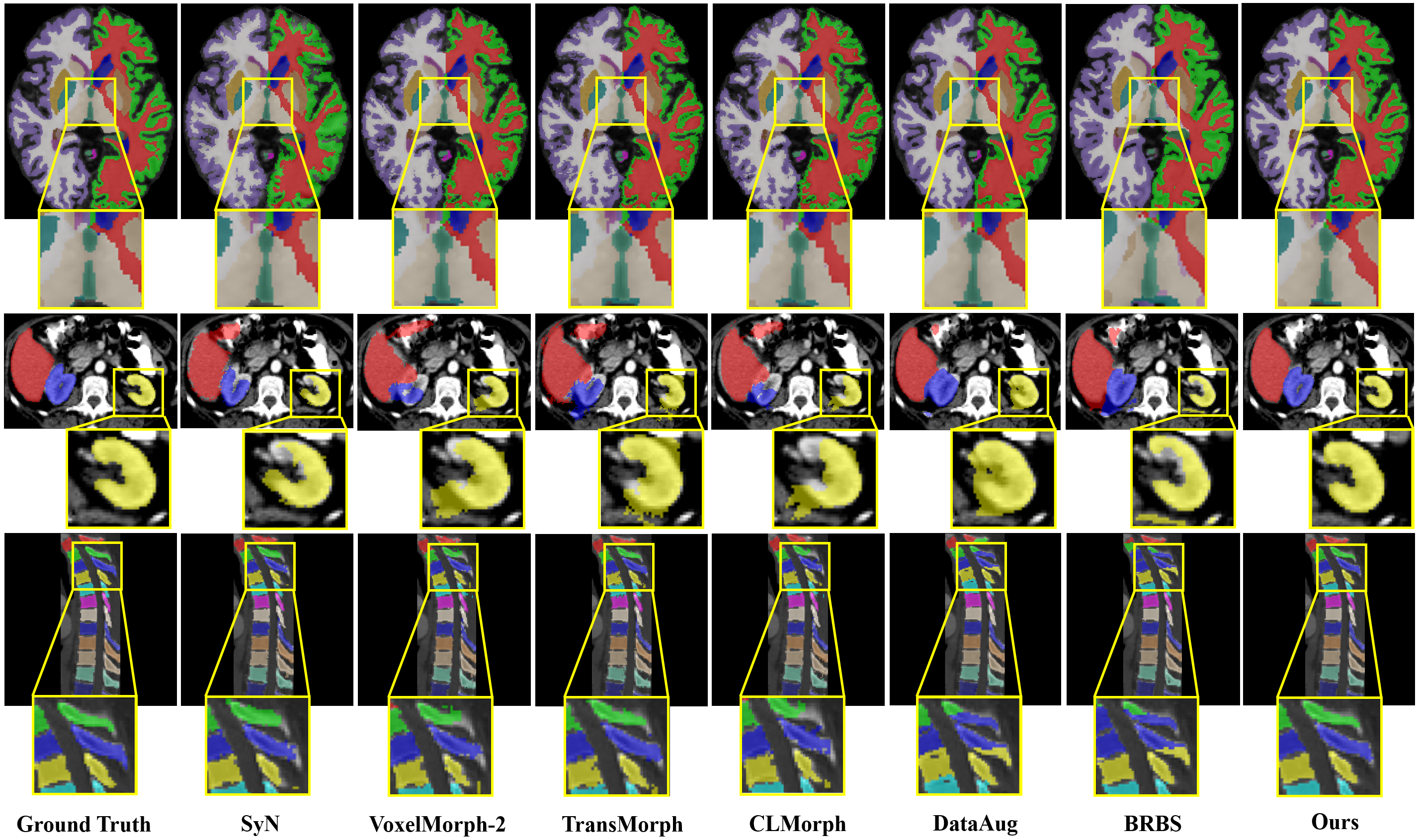}
    \caption{Comparison of our method with SOTA one-shot MIS methods on arbitrary cases. Yellow boxes indicate regions where our method is superior.}
    \label{fig:res_main}
\end{figure*}

\subsection{Implementation Details and Evaluation Metrics}
\label{Evaluation_Metrics}

The registration-based data augmentation network was trained for 500 epochs with a learning rate of $1\times10^{-4}$. Following \cite{liu2023contrastive}, we set hyperparameters $\alpha$ to 1 and $\beta$ to 0.01. Both teacher and student networks in the feature distillation module use 3D U-Net architectures with residual connections in the encoder and output segments. Despite similar structures, each network has unique output heads and trains for 200 epochs at a learning rate of $1\times10^{-3}$. We set hyperparameters $\lambda_{recon}=1$ and $\lambda_{hint}=1$ for OASIS brain and VerSe vertebrae segmentation tasks (see {Fig.} \ref{fig:hyperparam_analyse}a for the hyperparameters selection), and to 1 and 10 for the BCV abdominal organ segmentation task. Our method was implemented with PyTorch, optimized with Adam, and runs on NVIDIA RTX4090 GPUs with batch size of 2. Dice Similarity Coefficient (DSC) and the 95th percentile Hausdorff Distance (HD$_{95}$) were used to assess segmentation performance. A higher DSC is preferred, with 1 indicating perfect overlap and 0 indicating no overlap. A lower HD$_{95}$ is desirable, as it signifies better surface coincidence.



\subsection{Comparisons with Leading Methods}

In Table \ref{tab:three datasets res}, we compare our method with 12 others (Section \ref{Experimental_Settings}) on the OASIS, BCV, and VerSe datasets using DSC and HD$_{95}$mm metrics (Section \ref{Evaluation_Metrics}). Three key observations are: 1) LS methods show adaptability to one-shot segmentation but have poor segmentation quality (high HD$_{95}$) and unstable performance across labels. They perform well on large organs like the cerebral white matter but fail on smaller structures like vessels. 2) ABS methods perform stably for small structures due to consistent anatomical features. For OASIS dataset, they achieve nearly 80\% DSC, only 5.4\% lower than ours. However, their performance is limited by atlas-test image similarity and deformation extent, struggling with large deformations in BCV and VerSe tasks. Notably, the traditional ABS method SyN outperforms deep learning ABS methods in these tasks. 3) LRLS methods significantly outperform LS and ABS methods, thanks to deep learning and extensive training data, achieving over 80\% DSC across all tasks.

\begin{table*}[th]
\scriptsize
\setlength{\tabcolsep}{2pt}
\renewcommand{\arraystretch}{1.2}
        \caption{Comparison of segmentation performance for 35 brain structures on the OASIS dataset. Fully supervised serves as the upper bound benchmark. WM: white matter, CX: cortex, Vent: ventricle, STD: standard deviations. Scores for symmetrical brain regions are combined. The best result, excluding the upper limit, is highlighted in bold.}
    \label{tab:OASIS detail}
    \begin{tabular}{lccccccccccccccccccccc}

         ~ & ~ &\rotatebox{68}{Cerebral-WM} & \rotatebox{68}{Cerebral-CX} & \rotatebox{68}{Lateral-Vent} & \rotatebox{68}{Inf-Lat-Vent} & \rotatebox{68}{Cerebellum-WM}& \rotatebox{68}{Cerebellum-CX}& \rotatebox{68}{Thalamus} & \rotatebox{68}{Caudate} & \rotatebox{68}{Putamen} & \rotatebox{68}{Pallidum} & \rotatebox{68}{3rd-Vent} & \rotatebox{68}{4th-Vent} & \rotatebox{68}{Brain-Stem} & \rotatebox{68}{Hippocampus} & \rotatebox{68}{Amygdala} & \rotatebox{68}{Accumbens} & \rotatebox{68}{Ventral-DC} & \rotatebox{68}{Vessel} & \rotatebox{68}{Choroid-Plexus} & Mean$\pm$STD\\ 
        \toprule
       
        \textbf{Method} & \textbf{Type} &\multicolumn{20}{c}{\textbf{Dice Similarity Coefficient (DSC)\ $\uparrow$}}\\ 

        \midrule
        Supervised\cite{he2016deep}  &LS& \textit{0.970} & \textit{0.944} & \textit{0.957} & \textit{0.791} & \textit{0.959} & \textit{0.964} & \textit{0.961} & \textit{0.948} & \textit{0.951} & \textit{0.932} & \textit{0.923} & \textit{0.915} & \textit{0.969} & \textit{0.935} & \textit{0.923} & \textit{0.894} & \textit{0.929} & \textit{0.523} & \textit{0.764} & \textit{0.906$\pm$0.014}\\  
        \midrule
        U-Net\cite{ronneberger2015u} &LS&  
        0.900 & 0.862 &
        0.785& 0.033&
        0.865& 0.894&
        0.826& 0.752&
        0.841& 0.722&
        0.542& 0.717&
        0.909& 0.703&
        0.722& 0.473&
        0.611& 0.009&
        0.250& 0.684$\pm$0.064\\  
        ResUNet \cite{diakogiannis2020resunet} &LS & 
        0.940& 0.889&
        0.775& 0.087&
        0.867& 0.913&
        0.866& 0.804&
        0.866& 0.724&
        0.525& 0.773&
        0.900& 0.761&
        0.758& 0.563&
        0.775& 0.013&
        0.306& 0.724$\pm$0.048\\  
        \midrule
        Rigid\cite{arun1987least} &Trad & 0.638 & 0.521 & 0.620 & 0.181 & 0.720 & 0.785 & 0.792 & 0.627 & 0.744 & 0.710 & 0.619 & 0.574 & 0.840 & 0.627 & 0.685 & 0.517 & 0.728 & 
        0.156 & 0.199 & 0.597$\pm$0.049\\  
        Affine\cite{arun1987least} &Trad & 0.641 & 0.525 & 0.626 & 0.191 & 0.723 & 0.790 & 0.794 & 0.631 & 0.740 & 0.707 & 0.626 & 0.573 & 0.841 & 0.636 & 0.701 & 0.521 & 0.733 & 
        0.169 & 0.205 & 0.601$\pm$0.050\\ 
        SyN\cite{avants2008symmetric} &Trad & 0.821 & 0.702 & 0.894 & 0.359 & 0.842 & 0.888 & 0.916 & 0.862 & 0.881 & 0.891 & 0.855 & 0.816 & 0.939 & 0.804 & 0.842 & \textbf{0.828} & 0.880 & 0.458 & 0.525 & 0.784$\pm$0.021\\ 
        \midrule
        VoxelMorph-1\cite{balakrishnan2018unsupervised} &ABS& 0.868 & 0.738 & 0.889 & 0.472 & 0.865 & 0.893 & 0.913 & 0.848 & 0.894 & 0.892 & 0.836 & 0.801 & 0.937 & 0.806 & 0.830 & 0.793 & 0.878 & 0.469 & 0.527 & 0.793$\pm$0.019\\ 
        VoxelMorph-2\cite{balakrishnan2019voxelmorph} &ABS& 0.874 & 0.747 & 0.893 & 0.503 & 0.870 & 0.897 & 0.917 & 0.855 & 0.896 & 0.893 & 0.843 & 0.806 & 0.939 & 0.815 & 0.835 & 0.808 & 0.885 & 0.469 & 0.536 & 0.800$\pm$0.017\\ 
        TransMorph\cite{chen2022transmorph} &ABS& 0.866 & 0.725 & 0.882 & 0.492 & 0.863 & 0.887 & 0.919 & 0.841 & 0.894 & 0.893 & 0.832 & 0.794 & 0.937 & 0.813 & 0.839 & 0.816 & 0.883 & 0.476 & 0.535 & 0.796$\pm$0.017\\        
        CLMorph\cite{liu2023contrastive} &ABS& 0.875 & 0.748 & 0.890 & 0.498 & 0.872 & 0.898 & 0.916 & 0.849 & 0.896 & 0.891 & 0.838 & 0.802 & 0.939 & 0.811 & 0.832 & 0.803 & 0.881 & 0.477 & 0.534 & 0.799$\pm$0.019\\ 

        \midrule
        DataAug\cite{zhao2019data} &LRLS& 0.921 & 0.836 & 0.918 & 0.459 & 0.900 & 0.918 & 0.920 & 0.879 & 0.906 & \textbf{0.895} & \textbf{0.870} & 0.862 & \textbf{0.953} & 0.840 & 0.864 & 0.827 & 0.895 & \textbf{0.488}  & 0.565& 0.823$\pm$0.018\\ 
        BRBS\cite{he2022learning} &LRLS & 0.938 & 0.844 & 0.907  & 0.352 & 0.925 & 0.928 & \textbf{0.925} & 0.902 & 0.899 & 0.892 & 0.828 & 0.861 & 0.937 & 0.793 & 0.770 & 0.817 & 0.884 & 0.482 & \textbf{0.595}  &0.811$\pm$0.021\\ 
        \midrule
        \textbf{Ours} &LRLS & \textbf{0.952} &  \textbf{0.909} &  \textbf{0.925} & \textbf{0.641} & \textbf{0.935} & \textbf{0.937} & 0.919 & \textbf{0.915} & \textbf{0.917} & \textbf{0.895} & 0.836 & \textbf{0.888} & 0.951 & \textbf{0.878} & \textbf{0.867} & \textbf{0.828} & \textbf{0.894} & 0.485& 0.592 & \textbf{0.854$\pm$0.023}\\ 
    \bottomrule
    \end{tabular}

\end{table*}

Compared to other LRLS-based models, our method excels in two key areas. Firstly, it achieves the highest DSC and lowest HD$_{95}$ across all three datasets, demonstrating excellent performance and adaptability across different modalities and organs. Secondly, guided by distillation learning, it significantly enhances its understanding of anatomical features, leading to good segmentation performance for most organs. Segmentation details for each label in the OASIS (Table \ref{tab:OASIS detail}) and BCV (Fig. \ref{fig:bcv_boxplots}) datasets showcase the superior anatomical feature understanding of our framework.

As shown in Fig. \ref{fig:res_main}, our method demonstrates superior and robust performance across different organs and imaging modalities, notably in: 1) brain tissue segmentation: smoother cortical edges and more precise segmentation of structures like the thalamus and 3rd ventricle compared to methods like BRBS, which show misclassifications and coarser segmentation; 2) abdominal organ segmentation: accurate organ contours, unlike other methods with significant inaccuracies due to deformation registration challenges; 3) vertebrae segmentation: higher precision and clearer boundaries, whereas other methods show noticeable errors and blurred boundaries due to deformation limitations.

\subsection{Ablation Studies}
In this section, we evaluate the distinct impact of each main component in our proposed method, as well as the impact of integrated distillation learning and hyperparameters.
\subsubsection{Impact of Model Architecture Components}
Our model has three key components, as outlined in Table \ref{tab:ablation of network}. The ABS method (baseline) achieves a DSC of 79.9\% and an HD$_{95}$ of 3.214 mm. Adding the LRLS concept and training the student network with synthetic data (M2) slightly improves DSC by 0.7\%, but HD$_{95}$ increases by 0.611 mm. This limited improvement is due to the lack of teacher network guidance. With distillation learning (Ours), DSC improves with a 4.8\% and HD$_{95}$ decreases by 2.109 mm.

\begin{table}[ht]
\centering
\caption{Ablation study of model architecture components. SI+SL trains the student network with synthetic images and labels, while RI+SL uses real images with synthetic labels.}
\label{tab:ablation of network}
\setlength{\tabcolsep}{3pt} 
\renewcommand{\arraystretch}{1.1} 
\resizebox{0.48\textwidth}{!}{
\begin{tabular}{ccccccc}
\toprule
\textbf{Methods}&\textbf{Student} & \textbf{Teacher}  & \textbf{SI + SL}  & \textbf{RI + SL}  & \textbf{DSC $\uparrow$} & \textbf{HD$_{95}mm\downarrow$}\\
\midrule 
Baseline& & & & &  0.799$\pm$0.019 & 3.214$\pm$0.555  \\
M1&\checkmark & & & \checkmark&  0.788$\pm$0.016& 3.964$\pm$0.574 \\
M2&\checkmark & & \checkmark& &  0.806$\pm$0.020 & 3.825$\pm$2.077 \\
M3&\checkmark & \checkmark& & \checkmark& 0.817$\pm$0.018 &2.718$\pm$0.546 \\
\midrule 
\textbf{Ours}&\checkmark & \checkmark& \checkmark& &\textbf{0.854$\pm$0.023} &\textbf{1.716$\pm$0.252} \\
\bottomrule
\end{tabular}}
\end{table}

Additionally, Table \ref{tab:ablation of network} highlights the necessity of using synthetic images and labels (SI+SL) aligned via atlas distortion, enhancing learning efficiency. Conversely, using real images with synthetic labels (RI+SL) can introduce inaccuracies due to potential misalignment \cite{zhao2019data}. This is evident in M3, where RI+SL results in a 3.7\% decrease in DSC. Exposing the student network to real images does not necessarily improve performance. In M1, learning from RI+SL results in a DSC 1.8\% lower than M2 (SI+SL) and even lower than the baseline, underscoring the importance of alignment and the teacher network. Overall, the carefully designed components of our method are crucial for achieving the best performance.



\subsubsection{Impact of Integrated Distillation Learning}
\label{DL-impact}

Our feature distillation learning process (Eq. \eqref{kd-loss}) includes the segmentation loss \textbf{$\mathcal{L}_{seg}$}, the hint loss \textbf{$\mathcal{L}_{hint}$}, and the reconstruction loss \textbf{$\mathcal{L}_{recon}$}. Table \ref{tab:ablation of loss} demonstrates the effectiveness of each component. The results show that incorporating hint loss and reconstruction loss significantly enhances the student network's segmentation ability, indicating that our mix of loss functions ensures optimal segmentation results.

\begin{table}[ht]
\centering
\caption{Impact of integrated distillation learning module.}
\label{tab:ablation of loss}

\begin{tabular}{cccccc}
\toprule
\textbf{$\mathcal{L}_{seg}$} & \textbf{$\mathcal{L}_{hint}$} & \textbf{$\mathcal{L}_{recon}$ } & \textbf{DSC $\uparrow$} & \textbf{HD$_{95}mm\downarrow$}\\
\midrule 
 &\checkmark & \checkmark &  0.003$\pm$0.000 & -  \\
\checkmark& & \checkmark &  0.806$\pm$0.020 & 3.825$\pm$2.077\\
\checkmark& \checkmark& &  0.826$\pm$0.023 & 2.973$\pm$1.299\\
\checkmark& \checkmark & \checkmark & \textbf{0.854$\pm$0.023} & 
 \textbf{1.716$\pm$0.252}\\
\bottomrule
\end{tabular}
\end{table} 

\begin{figure*}[ht]
    \centering
    \includegraphics[width=0.95\textwidth]{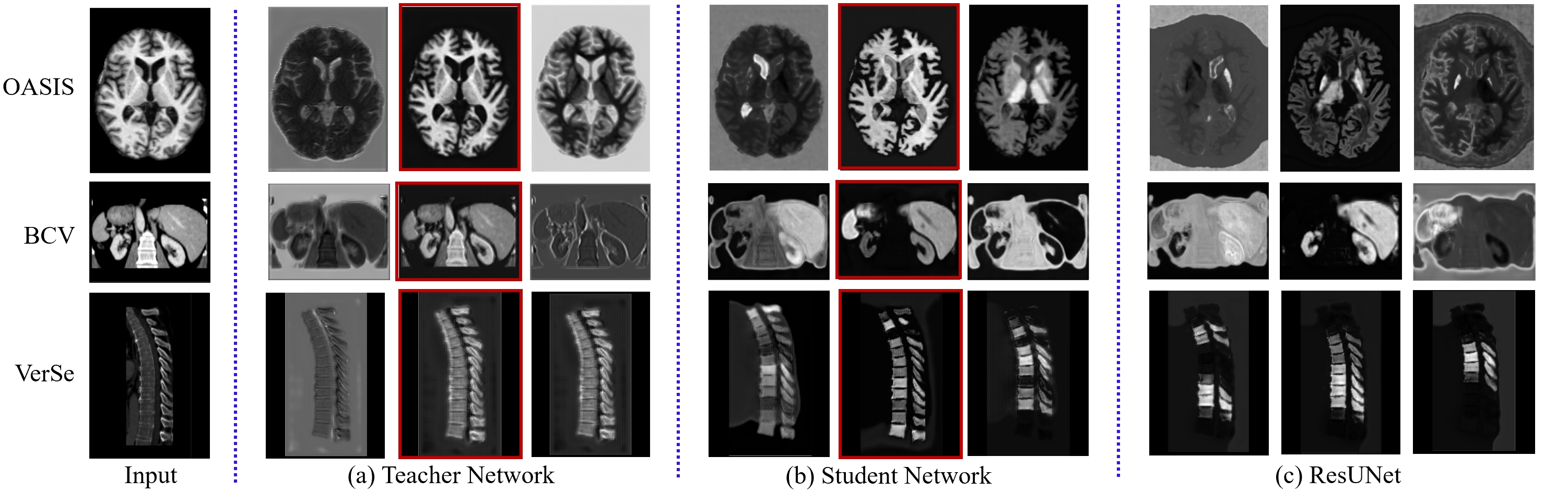}
    \caption{Feature maps from the last layer of each network for three randomly selected channels of an arbitrary sample in the OASIS, BCV, and VerSe datasets. ResUNet is the student network without teacher guidance. Red boxes highlight feature maps of the student network that closely resemble the teacher's.
    }
    \label{fig:features-3}
\end{figure*}

To further assess the significance of our proposed hint loss (Eq. \eqref{hint-loss}), we conducted an ablation study to examine the segmentation performance using \textbf{$\mathcal{L}_{hint}$} with L2 norm and cosine similarity. As shown in Table \ref{tab:ablation of hint loss}, employing cosine similarity as the hint loss outperforms L2 loss within the distillation learning framework. These findings confirm that using cosine similarity reduces the feature distance, thereby enhancing the segmentation performance of the student network. Fig. \ref{fig:features-3} shows the necessity of our designed teacher-student network by comparing the feature maps for all three datasets. For accurate comparisons, all compared models share the same network architecture.

\begin{table}[h]
\caption{Comparison of different hint losses in distillation learning.}
\label{tab:ablation of hint loss}
\centering
\begin{tabular}{l|cc}
\toprule
\textbf{Methods} & \textbf{DSC $\uparrow$} &\textbf{HD$_{95}mm\downarrow$}  \\ 
\midrule 
L2 loss & 0.811$\pm$0.017 & 2.992$\pm$0.612\\ 
Cosine similarity loss & \textbf{0.854$\pm$0.023} &\textbf{1.716$\pm$0.252}\\ 
\bottomrule
\end{tabular}
\end{table}

\subsubsection{Impact of Hyperparameters}
\label{hyper-impact}

\begin{figure}[ht]
    \centering
    \includegraphics[width=0.5\textwidth]{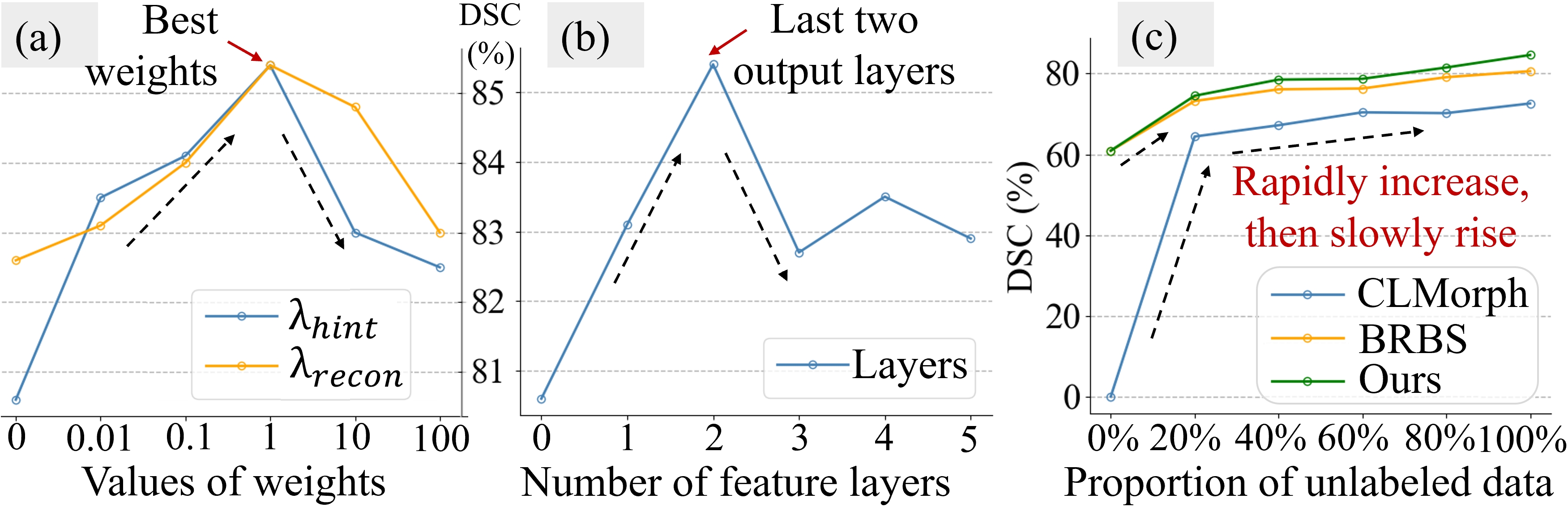}
    \caption{Hyperparameters analysis: (a) Examining weights for reconstructive loss and hint loss in Eq. \eqref{kd-loss}, (b) choosing the number of feature output layers for distillation learning on the OASIS dataset, and (c) determining the size of unlabeled data on the BCV dataset.}    
    \label{fig:hyperparam_analyse}
\end{figure}

We first performed hyperparameter tuning to find the optimal weights of ${\lambda}_{recon}$ and ${\lambda}_{hint}$ in our loss function (Eq. \eqref{kd-loss}). As shown in Fig. \ref{fig:hyperparam_analyse}a, increasing ${\lambda}_{recon}$ and ${\lambda}_{hint}$ from 0 to 1 enhances the model's DSC, with optimal performance at 1. Beyond 1, performance declines, indicating excessive weights negatively affect $\mathcal{L}_{seg}$.


In Fig. \ref{fig:hyperparam_analyse}b, our analysis shows that using the output features from the last two layers for calculating the hint loss $\mathcal{L}_{hint}$ yields the best distillation learning outcomes.


Next, we evaluated the impact of unlabeled data on our method compared to other leading one-shot MIS methods. As shown in Fig. \ref{fig:hyperparam_analyse}c, our segmentation performance improves rapidly with up to 20\% unlabeled data, then slows as more unlabeled data is added. This is because our teacher network effectively guides the student network with high-quality features from real data. On the BCV dataset, even using 100\% of the unlabeled images amounts to only 29 images, showing good performance with minimal unlabeled data. In contrast, ABS methods like CLMorph rely heavily on unlabeled data and cannot function without it, highlighting their limitations. On the other hand, our method and BRBS degrade to LS methods when unlabeled data is absent, using only the supervised model for one-shot segmentation. Our method consistently achieves higher DSC than BRBS, demonstrating the robustness of our framework regardless of the amount of unlabeled data.

\section{Discussion}
ABS methods \cite{arun1987least,avants2008symmetric,balakrishnan2019voxelmorph,balakrishnan2018unsupervised,chen2022transmorph,liu2023contrastive} generate predicted labels by propagating atlas labels through registration. Their segmentation accuracy tends to unstable due to reliance on the similarity between atlas and target images, and they lack robustness for segmenting large organs with significant deformations, as shown in abdominal CT (Figs. \ref{fig:bcv_boxplots}-\ref{fig:res_main}). These findings are consistent with previous literature \cite{van20213d}.

In our experiments, the LRLS methods \cite{zhao2019data,he2022learning} consistently outperformed ABS methods in segmentation tasks. Current LRLS methods either synthesize richer pseudo-datasets through carefully designed sampling strategies \cite{zhao2019data,ding2021modeling,he2022learning} or enhance performance under small sample conditions via joint registration-segmentation optimization \cite{he2022learning,he2020deep,xu2019deepatlas}. Despite their effectiveness, these approaches often limit the quality and quantity of synthetic images when faced with challenging or scarce unlabeled data. Joint optimization models are parameter-heavy, difficult to train, and have limited adaptability to different modalities and organs. Additionally, they fail to fully utilize real unlabeled images, which contain more reliable anatomical structures than synthetic images.

Based on the LRLS paradigm, we introduced a novel one-shot MIS framework that allows networks to directly ‘see’ real images through a distillation learning process guided by image reconstruction. As seen in Fig. \ref{fig:features-3}, the teacher network generates clear, detailed feature maps, providing high-quality feature representations. The student network, guided by the teacher, learns these features and shows higher activation in regions of interest, ensuring segmentation accuracy. The student network's feature maps retain most of the critical details from the teacher network, and some are very similar (highlighted by red boxes), indicating effective guidance. In contrast, feature maps from ResUNet without teacher guidance are blurry and lack detail. For example, in vertebrae segmentation, the student network's feature maps clearly outline various organs, effectively distinguishing foreground from background. Without the teacher's guidance, ResUNet's highly activated regions become blurry and incomplete, showing that an isolated student network cannot capture critical features, thus affecting segmentation performance.

While our method demonstrated promising results, it has several limitations. Medical images are highly heterogeneous, and labeling anatomical structures is both costly and time-intensive. As with other one-shot MIS methods, our framework is currently tailored for scenarios with limited labeled samples. This constraint may hinder the model's ability to capture complex anatomical structures and diverse imaging features, potentially leading to performance discrepancies between training data and unseen test data from unknown domains. Future work will address these challenges by evaluating the generalizability of our method on larger datasets with more comprehensive annotations. For instance, the proposed method could be extended to segment various organs (e.g., brain, abdomen and vertebrae) for diverse clinical applications. Specifically, we are actively working on applying the framework to a large dataset of pediatric CT images to segment whole brain structures, aiming to support the assessment of brain diseases and developmental processes.

\section{Conclusion}
In conclusion, we present a groundbreaking one-shot medical image segmentation framework with four key advancements: 1) An unsupervised teacher network that reconstructs real images to guide the student network's training on synthetic data; 2) An advanced feature distillation framework for precise segmentation; 3) A streamlined and efficient inference network; 4) Superior performance over state-of-the-art methods across three datasets with different organs (brain, abdomen, vertebrae) in MRI and CT modalities. Our method shows exceptional generalizability and potential for improving diagnostic and treatment accuracy for multiple diseases.


\bibliography{ref}
\bibliographystyle{ieeetr}

\end{document}